\documentclass[aps,pra,twocolumn,amssymb,amsmath,color]{revtex4}
\usepackage{amsmath}
\usepackage{amssymb}
\usepackage{latexsym}
\usepackage{revsymb}
\usepackage{epsfig}
\begin{document}
\title{On conservation laws in quantum mechanics}

\author{Yakir Aharonov$^{a,b}$}
\author{Sandu Popescu$^{c}$}
\author{Daniel Rohrlich$^d$ }
\affiliation{$^a$School of Physics and Astronomy, Tel Aviv University, Tel Aviv 69978 , Israel.}
\affiliation{$^b$Department of Physics, Chapman University, Orange CA, USA.}
\affiliation{$^c$H. H. Wills Physics Laboratory, University of
Bristol, Tyndall Avenue, Bristol BS8 1TL}

\affiliation{$^d$Physics Department, Ben-Gurion University of the Negev, Beersheba 8410501, Israel.}

\begin{abstract}
 We raise fundamental questions about the very meaning of conservation laws in quantum mechanics and we argue that the standard way of defining conservation laws, while perfectly valid as far as it goes, misses essential features of nature and has to be revisited and extended.

\end{abstract}


\maketitle

\newcommand{\tr}{\mbox{Tr}}
\newcommand{\id}{\mbox{\bf I}}
\newcommand{\I}{\mbox{\bf i}}
\newcommand{\ket}[1]{\left | #1 \right \rangle}
\newcommand{\bra}[1]{\left \langle #1 \right |}
\newcommand{\braket}[2]{\left \langle #1 | #2 \right \rangle}
\newcommand{\up}{\uparrow}
\newcommand{\down}{\downarrow}
\newcommand{\beqa}{\begin{eqnarray}}
\newcommand{\eeqa}{\end{eqnarray}}
\newcommand{\beq}{\begin{equation}}
\newcommand{\eeq}{\end{equation}}
\newcommand{\ra}{\rangle}
\newcommand{\la}{\langle}

\section{Introduction}

Conservation laws, such as those for energy, momentum, and angular momentum, are among the most fundamental laws of nature. As such they have been intensively studied and extensively applied. First discovered in classical Newtonian mechanics, they are at the core of all subsequent physical theories, non-relativistic and relativistic, classical and quantum. Here we present a paradoxical situation in which such quantities are seemingly not conserved. Our results raise fundamental questions about the very meaning of conservation laws in quantum mechanics and we argue that the standard way of defining conservation laws, while perfectly valid as far as it goes, misses essential features of nature and has to be revisited and extended.

That paradoxical processes
must arise in quantum mechanics in connection with conservation laws is to be expected. Indeed, on the one hand, physics is local:
causes and observable effects must be locally
related, in the sense that no observations in a given space-time region can yield any information about events that take place outside its past light cone \cite{footnote}.
On the other hand, measurable dynamical quantities are identified with eigenvalues of
operators and their corresponding eigenfunctions are not, in general,
localized.  Energy, for example, is a property of an entire wave function.  However, the law of conservation of energy is often applied to
processes in which a
system with an extended wave function interacts with a
local probe.
How can the local probe ``see" an extended wave function?  What determines
the change in energy of the local probe?  These questions lead us to
uncover quantum processes that
seem, paradoxically, not to conserve energy.

The present paper (which is based on a series of unpublished results, first described in  \cite{old} and \cite{thesis}), presents the paradox and discusses various ways to think of conservation laws but does not offer a resolution of the paradox. A subsequent paper will present our resolution. The reason for publishing the paradox and resolution separately is that the paradoxical effect stands alone - its existence is independent of attempts to explain it - while readers may disagree with our proposed resolution.

\section{Superoscillations}

Essential to this paper is a mathematical structure we call ``superoscillation". Common wisdom assumes that no function can oscillate faster than its fastest Fourier component. Yet as we show here, there is a large class of functions for which this assumption fails. Indeed we have found functions that oscillate, on a given interval, arbitrarily faster than the fastest Fourier component. An example of such a function is the following:

\begin{equation}f(x)=\left( {{1+\alpha}\over 2} e^{i{{x}/N}}+{{1-\alpha}\over 2} e^{-i{{x}/ N}}\right)^N~~~,\label{state}\end{equation}
where $\alpha$ is a positive real number, $|x|\leq \pi N$,  and $N$ is a large integer. An extensive discussion of the properties of this function, first introduced in \cite{old,thesis}, appears in \cite{berry1, berry2, daniele}. To display its basic properties, we first write it, via the binomial formula, as

\begin{equation}f(x)=\sum_{n=0}^N c(n;N,\alpha)e^{i {{(2n/N-1)}}x}~~~,\label{sumoverfrequencies}\end{equation}
where the $c(n;N,\alpha)$ are constants;  \begin{equation}c(n;N,\alpha)={1\over{2^N}}\left(\begin{matrix} N\\ n\end{matrix}\right)(1+\alpha)^n(1-\alpha)^{N-n}.\end{equation}
From (\ref{sumoverfrequencies}) one can see that $f(x)$ is a sum over wave numbers $k_n={{2n/N-1}}$, ranging from -1 to 1.

Now consider this function in the region $|x|\lesssim\sqrt N$. Here we can approximate the exponentials by their first order Taylor expansion and obtain

\beqa &&f(x)\approx\left( {{1+\alpha}\over 2} (1+i{{x}\over N})+{{1-\alpha}\over 2} (1-i{{x}\over N})\right)^N\nonumber\\ & \quad &=\left(1+{{i\alpha x}\over N}\right)^N
\approx e^{i\alpha x}\label{superoscillations}.\eeqa

\bigskip
\noindent
Hence, in the restricted region, $f(x)$ behaves as an oscillation of wave number $\alpha$. But, crucially, $\alpha$ need not be smaller than 1. By taking  $\alpha\gg 1$, we ensure that in the region of validity of the approximation, $|x|\lesssim\sqrt N$, the function $f(x)$ oscillates with wave number $\alpha\gg 1$ although all its Fourier components have wave number smaller than 1.  In other words, a superposition of long wavelengths, the longest being  $2\pi N$ and the shortest being $2\pi$, can, in the region $|x|\lesssim\sqrt N$, oscillate with the much shorter wavelength ${{2\pi}/{\alpha}}$. Furthermore, the region of these ``superoscilations" can be made arbitrarily large and include arbitrarily many wavelengths by taking $N$ sufficiently large.

Note that there is no contradiction with the Fourier theorem since the region where this function is (almost) identical to an oscillation of a frequency not contained by its Fourier decomposition does not extend over the entire region where the function is defined.

Although it is not essential for our present paper, it is interesting to note that outside the superoscillatory region $f(x)$ increases exponentially. This is a generic property of functions with superoscillatory regions.

\section{The experiment}

\bigskip
\noindent
The type of effect we describe here is common for all conserved quantities that depend on the shape of the wave function all over the space including energy, momentum and angular momentum. Here we will focus on energy, for which the proof is more intuitive.

\bigskip
\noindent
Consider a box of length $2\pi Na$ (where $a$ is some unit length) that contains a single photon in the state $\psi(x)$
\beq\psi(x)={i\over{{\cal N}}}\big( f(x/a)-f^*(x/a)\big)\label{particle state}\eeq
with $f(x)$ defined in Eq. (\ref{superoscillations}) and $\alpha\gg 1$. Here ${\cal N}$ is a normalisation factor.
$\psi(x)$ has properties similar to $f$ but it obeys the boundary conditions $\psi(-\pi Na)=\psi(\pi Na)=0$ at the walls of the box. (See Fig. 1.)

\begin{figure} [h]
\epsfig{file=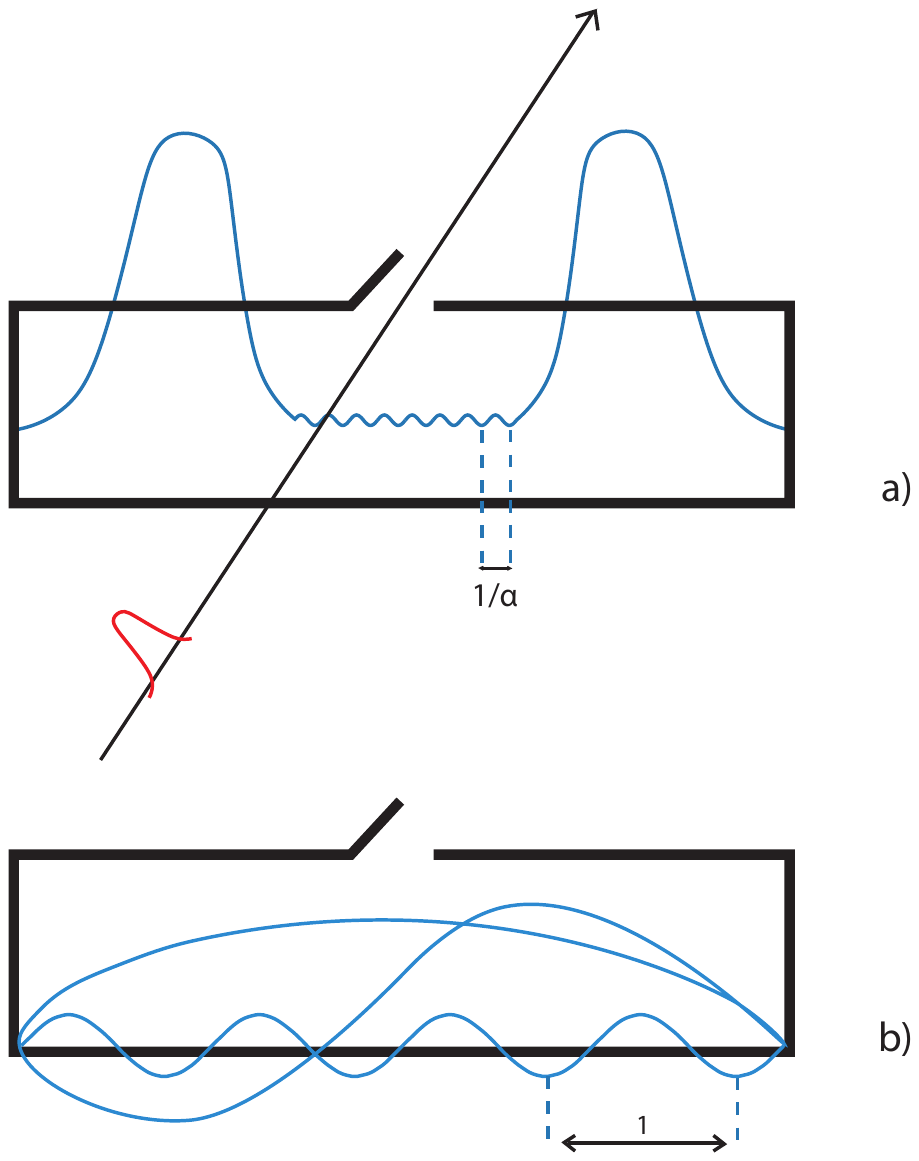, scale=0.50} \caption{(a) A photon in the box in the specially prepared low-energy quantum state which, in the central region, oscillates with a spatial frequency greater than that of any Fourier components (see (b)). As the opener is passing by, it opens the box and extracts the photon, if the photon is there. The shape of the wave function is not accurate but simply illustrative; the true wave function is exponentially larger away from the center and has a more complicated shape. }
\end{figure}

From now on however, for simplicity, we take $a=1$ and we work in the usual units $\hbar=c=1$.

Given the relation between wavelength, frequency and energy for the photon, the decomposition (\ref{sumoverfrequencies}) shows that the photon is in a superposition of different energy eigenstates with wave numbers
\beq k_n=(2n/N-1)\eeq all smaller than or equal to  $1$ (in absolute value), corresponding to energy
eigenvalues

\begin{equation}
E_n=|k_n|
\end{equation}
with the maximal energy $E_{max}=1$. On the other hand, we also know that in the region $|x|\lesssim\sqrt N $ around the center of the box, the wave function of the photon resembles that of a monochromatic photon with wave number $\alpha$ hence of energy

\beq{\cal E}=\alpha>>E_{max}=1.\eeq In other words, in the box we have a low-energy photon, which in the center of the box looks like a high-energy photon.

\bigskip
\noindent
Suppose now that a mechanism that we will call the ``opener" opens the box in the center for a time $T\lesssim{{\sqrt N}}$ and inserts a mirror, such that if the photon hits the mirror, it comes out of the box (as in Fig. 1). This happens if the photon is situated at a distance not larger than $T$ from the mirror; otherwise it cannot get there while the box is open. The probability for this to happen is at most
$\int_{-T}^{T}|\psi (x)|^2 dx\label{probability}$.

\bigskip
\noindent
Now suppose we find the photon out of the box.  What is its energy?

\bigskip
\noindent
Naively we would think that the emerging photon must have one of the  energies  $ E_n=|{2n/N-1}|\leq 1$ that it had originally in the box - after all, reflection from a mirror doesn't change the spectrum of light. On second thought, however, we realize that this cannot be so. Indeed, the box was open only for a time $T$. In the entire region $|x|\leq T\leq \sqrt N$ around the opening, the wave function of the photon was essentially indistinguishable from a plane wave of high energy $\cal E$. The information revealing that the photon was {\it not} a true monochromatic photon of this high energy is contained in the shape of the wave function in regions situated farther from the opening than $T$; relativistic causality dictates that this information could not arrive at the opening during the time it was open. Hence the photon that emerges from the box must be in a state which is identical to that in which a genuine photon of energy $\cal E$ would emerge from the box. But for this second case it is trivial to see what happens:  the mirror just reflects the photon out of the box but it doesn't change it frequency and energy.

In addition, the wave function is chopped into a wave-train of length $T$ by our closing the box after time $T$. Hence if a genuine photon of energy ${\cal E}=\alpha$ emerged from the box its energy spectrum would have a peak at energy $\alpha $ and a spread in energy of ${1/T}$. By increasing $N$ we can increase $T$ and hence reduce by as much as we want the disturbance produced by the finite opening time of box; the photon thus emerges as close as we want to its initial energy ${\cal E}$.  We thus conclude that when our ``fake high-energy" photon emerges from the box it must have energy ${\cal E}$ exactly as the genuine high-energy photon and not the low energies it originally had; see Figs. 2 and 3.

\begin{figure} [h]
\epsfig{file=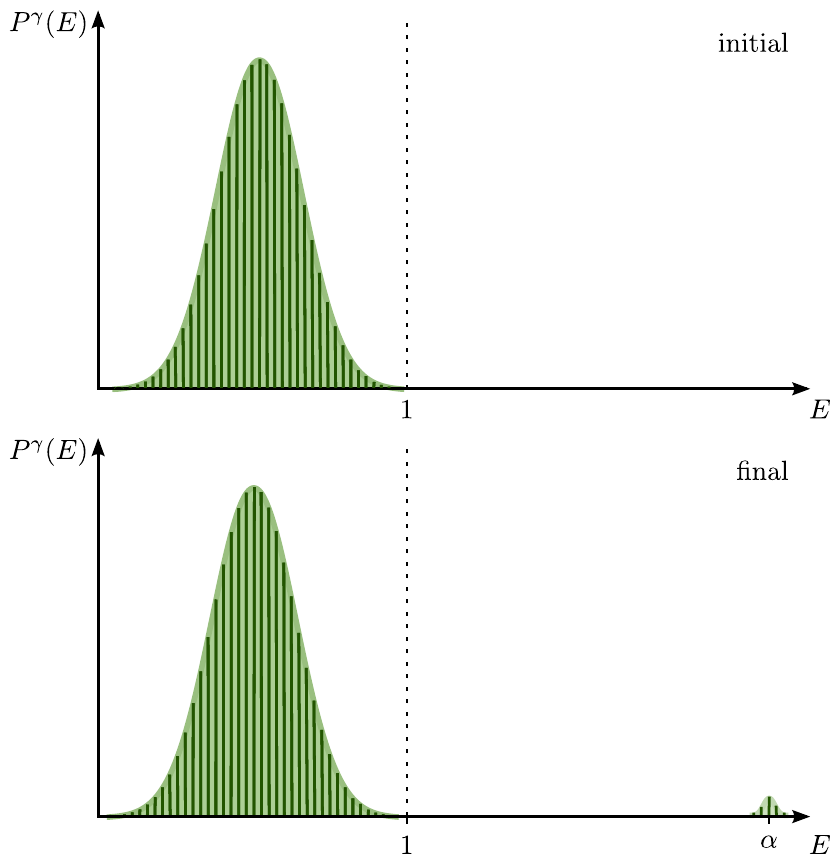, scale=0.50} \caption{The initial and final distributions of the energy of the photon. Initially it was a superposition of low energies and strictly no energy higher than 1. Finally a peak at energy $\alpha$ appears, which corresponds to the extracted photons. }
\end{figure}
\begin{figure} [h]
\epsfig{file=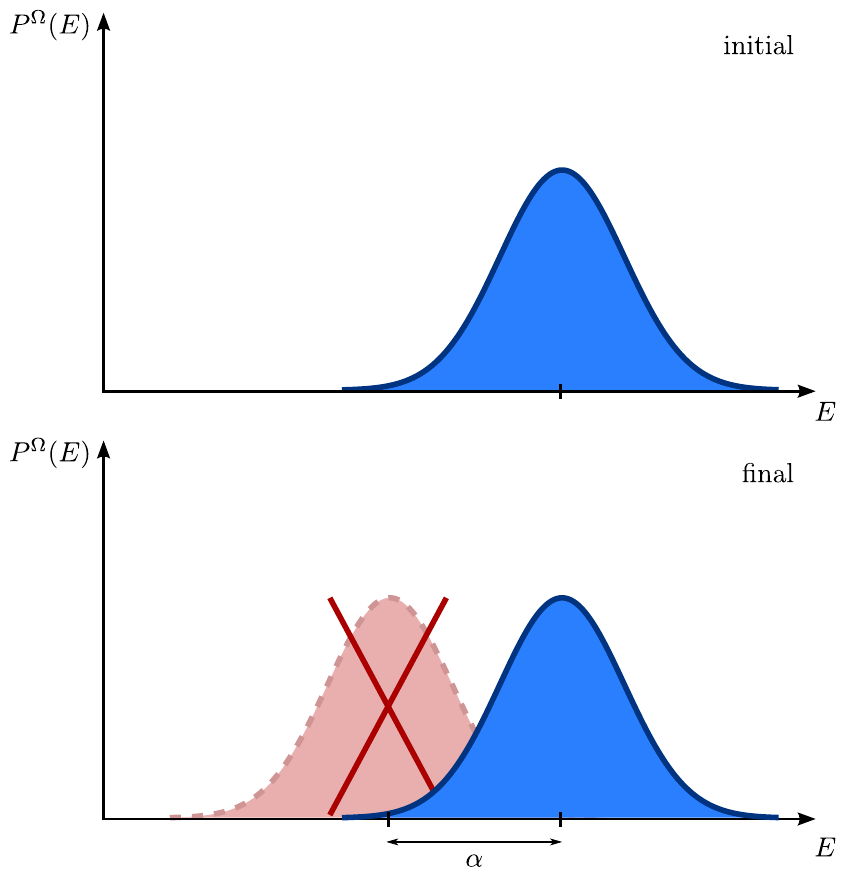, scale=0.50} \caption{The initial and final distribution of the energy of the opener for the cases when the photon emerges from the box. One would expect the final energy of the opener to be lower than the initial, to compensate for the increase of the energy of the photon, but this does not happen; instead the final energy distribution is the same as the initial (up to small perturbations due to the truncation of the photon wave packet). }
\end{figure}

To summarize, when the photon emerges from the box, it has much higher energy than it initially had. Where did the extra energy come from? This is the question that concerns us in this paper.

\section{The paradox}

Inside the box the photon was in a superposition of different energies eigenstates, all smaller than 1, and out it emerges with the much higher energy $\alpha$. Where does the extra energy come from? A first guess is that the energy comes from the mechanism used for extracting the photon. Indeed, we need to open the box, insert the mirror, then take out the mirror and close the box. This mechanism, which we call ``opener", effectively subjects the photon to a time-dependent Hamiltonian, and such a Hamiltonian need not conserve energy.

To put it differently, we can look at the total Hamiltonian, describing the photon and the opener. The total Hamiltonian is time independent since it describes the total system, with no parts left outside; the time dependence seen by the photon comes from the time evolution of the opener and its interaction with the photon. The total energy is  conserved for time-independent Hamiltonians.  Thus, we are tempted to say, all that happens is that the opener and the photon exchange energy: when the photon emerges from the box, and, as we proved, has higher energy than it had inside the box, the opener must have lost the same amount - a trivial case of energy exchange.

We now arrive at the crux of the problem. Although this explanation is  the most natural, it is wrong: the photon could not have gotten its energy from the opener. The reason is again causality, as we shall now see.

Consider the case of a monochromatic high-energy photon of energy $\alpha$. When this photon emerges from the box its energy is unchanged (up to fluctuations of order $1/T$). Hence in this case  the opener does not give it any energy. But then the opener cannot give energy to the fake photon either.

Indeed, recall that the entire difference between the high-energy photon of energy $\alpha$ and our specially prepared low-energy photon lies in regions situated further from the center than $\sqrt N$; this information cannot arrive at the opener during the time of the experiment, $T\lesssim {\sqrt N }$. Immediately after the experiment is over, we can measure the energy of the opener.  Since the opener is localized in a small region (around the center of the box) we have immediate access to it and can measure its energy in a short time; the measurement can be finished long before information from $|x|\geq \sqrt N $ can arrive. If  by measuring the opener we could determine  whether the box contained the original low energy photon or the monochromatic high-energy photon, we would violate relativistic causality. Thus, since the opener didn't lose energy when the box contained a high-energy photon, it cannot lose energy in the case of the low-energy photon either! We must therefore conclude that the extra energy did not come from the opener.

This is the paradox. The photon emerged from the box with energy much higher than it had inside, but the energy did not come from the opener, the only other system in the problem. Energy seems not to be conserved.

\section {Energy conservation}

Faced with this paradox one can respond in various ways. The conventional response is that there is problem whatsoever, and there cannot ever be. In quantum mechanics, the standard formulation of a conservation law is that the {\it probability distribution}  of the conserved variable over the {\it entire ensemble} should not change. This law applies to {\it any} time-independent Hamiltonian. On this basis, there should be absolutely no energy non-conservation in our example either.  And, of course, from this point of view, there is none. Indeed, in the preceding sections we focused on what happens when the photon emerges from the box. But it is also possible (and actually far more probable) that the photon does not emerge from the box. It happens, because the wave function of the photon extends all over the box, so the photon has a non-zero (and in fact quite large) probability to be far from the central region. If so, it cannot reach the opening while the box is open therefore it cannot leave the box. To see standard energy conservation at work, we must consider these cases as well. What we find in these cases is that again the opener didn't lose any energy (since it did not collide with the photon), but the photon remains in the box with lower energy (as the wave function loses its superoscillatory piece) - again a paradox. Considering these cases as well we find that, as expected, the probability distribution of the {\it total} energy (photon plus opener) did not change.

But - and this is the main point of our paper - we would like to argue that the standard formulation of conservation laws, though absolutely correct as far as it goes, is simply not enough. The standard conservation law is statistical and says nothing about individual cases. We would like to argue, however, that it is legitimate to  ask what happened in a particular individual case. In our example, suppose we have in the box a photon of energy of order 1 eV (more precisely a photon in a superposition of various energies but absolutely none of them larger  than 1 eV).  Yet, when we open the box the photon emerges with energy of order of 1 GeV. We should definitely be entitled to ask where the energy came from.

What we showed in our example is that this energy cannot come from the mechanism that extracts the photon from the box. Since this is the single other system in the problem, we are faced with energy non-conservation in this individual case. So we do have a problem that needs to be explained.

Furthermore, we also argue that the standard formulation is not only limited in that it cannot address individual cases, it is also unsatisfactory in the meaning of the story it tells.

Suppose that we repeat our experiment a large number of times. Consider a large number of boxes, each box containing just one single photon, prepared in the special low energy state $\psi(x)$ of Eq. (\ref{particle state}), each box having its associated opener. In some experiments the photon comes out of the box; in others the photon remains inside. In each case the energy of the opener remains unchanged (apart from fluctuations of the order ${1/T}<<{\cal E}-E_{max}$). Since the photons that emerge from their boxes have increased energy, for the average energy to remain constant the photons that stay in their boxes must be left with lower energy. But what this effectively means is that the photons that emerged from their boxes got their energy from the photons in the other boxes. But the experiments are completely independent; they may even happen at different times and in different places.  Nevertheless, the
photons which stay inside their boxes supply energy to the ones that emerge from the other boxes! Clearly the idea is absurd and it cannot be an acceptable resolution of our paradox.

Another possible response to the paradox is to argue that it makes no sense to talk about the energy of a photon as long as it is in a superposition of different energy eigenstates. However we note that the photon had {\it zero} probability to have any energy larger than $E_{max}=1$, yet it emerges with the high energy ${\cal E}>>E_{max}$. So the paradox of the photon's extra energy remains.

As noted in the introduction, we do not offer any resolution here; we leave the paradox open. But below we provide more details. First we present an explicit model; then we analyze in more detail the standard energy conservation as applied to our situation. Although there are no contradictions here, the specific way in which the energy is conserved in the statistical ensemble is extremely unusual and instructive.

\section {Explicit model}

We now give an explicit model. Since relativistic quantum field theory has  well-known technical difficulties,  we will formulate a non-relativistic model. The experiment is the same, the only difference being that instead of a photon, the box contains a non-relativistic particle. Of course, we are now no longer allowed to use relativistic causality arguments, and we will prove our statements  by explicit calculations. Nevertheless, the intuition for the non-relativistic model is exactly the same as in the relativistic case since also non-relativistic quantum mechanics allows finite time intervals in which a part of a wave function can act, to a good approximation, independently of the rest \cite{robinson-lieb}.

Consider the following
Hamiltonian to model our experiment:
\begin{equation}H =
{{p^2}\over{2}}+
V(x) {1+\sigma_z \over 2}+ p_q + {{\pi}\over 2}g(x) \delta (q) \sigma_x .\end{equation}
The low-energy particle in the box has coordinate $x$, momentum $p$ and mass $m=1$, while the ``opener" is modeled by a particle with coordinate $q$ and momentum $p_q$.  $V(x)$ is an infinite square potential well which represents the box: it is zero inside the box ( i.e. for $|x| \le \pi N$) and it is infinite outside.

We let our particle have  an internal degree of freedom, a ``spin", which determines whether the particle is in the box or free. The states $|\uparrow \rangle$ and $|\downarrow \rangle$ are eigenstates of $\sigma_z$. When the spin is $|\uparrow \rangle$ the potential $V(x)$ confines the particle inside the box; when the spin is $|\downarrow \rangle$ the particle is free since the term in $1 + \sigma_z$ multiplying $V(x)$ vanishes.

The opener's free hamiltonian is $p_q$ while the last term in $H$ describes the particle-opener interaction. The interaction takes place when the opener is at $q=0$; at all other times the opener is free. The opener moves at constant speed $\dot q=1$ without spreading, both when it is free and while the interaction takes place. The opener moves from $q<0$ where it is free, through the interaction region, $q=0$, to $q>0$ where it is free again. (One may recognize the opener as the model of an ideal clock that turns on and off an interaction when the ``clock time" indicated by the pointer $q$ is $q=0$.)

The interaction term is designed to release the particle if it is situated in a window around the center of the box. This works as follows. The operator $\sigma_x$ can release the trapped particle by flipping $|\uparrow \rangle$  to $|\downarrow \rangle$.  But the particle is released only if it is situated in a window around the center of the box. The window
is determined by $g(x)$, which is zero outside a window $|x|\le L \lesssim\sqrt N$ and $g(x)=1$ inside. The interaction term is thus non-zero only when the particle is in this window, hence only if it is here can the particle be released. (Note that this toy model differs slightly from the example in the previous sections: There the window was taken to be small, but open for a time long enough for a wave-train of length $L$ to emerge through it. Here the window is open for an infinitesimal time, but is large enough to let a wave-train of length $L$ emerge from it.)

Consider now that at time $t=0$,
we prepare the trapped particle in a state
$\Psi (x,0) |\uparrow \rangle$ with $\Psi(x,0)$ equal to our special state $\psi(x)$ given by Eq. (\ref{particle state}). In the region $|x|\lesssim\sqrt N $ the wave function $\Psi(x, 0)$ looks like a high-energy state. In relativistic quantum mechanics, when the particle is situated in this region its time evolution is identical to that of a high-energy particle, since the information that this is not a true high-energy particle is contained only in faraway regions of space and it takes a finite time to arrive in the center of the box. But, as we mentioned above, the same is (approximately) true also in non-relativistic quantum mechanics (see Supplementary Information 1). Therefore for $|x|\lesssim\sqrt N $ the time evolution of the particle (in the absence of the interaction with the opener) is $\Psi(x,t)=e^{-i\alpha^2t/2m}\Psi(x,0)$, i.e. it just accumulates the same time dependent phase as the bona-fide energy eigenstate $\sin(\alpha x)$. This approximation is valid for any time $T\lesssim\sqrt N $. It is during this time that we extract the particle from the box.

To ensure that the opener opens the box during this time window we choose its initial state
$\phi (q)$ to have support only for $q$ within
$-T\le q \le 0$.

We have now to calculate the time evolution of the particle-opener system
\beq e^{-i{H}T} \phi (q) \psi (x) |\uparrow \rangle=\Theta_{\uparrow}(q,x)|\uparrow \rangle+\Theta_{\downarrow}(q,x)|\downarrow \rangle.\label{final_state}\eeq
As noted above, if the particle is found in the central part of the box, after the interaction with the opener it gets released; otherwise it remains inside the box. This corresponds to the final state (\ref{final_state}) being a superposition of two terms, one with the ``spin" $\downarrow$ and one $\uparrow$ respectively. (Note that $\Theta_{\uparrow}$ and $\Theta_{\downarrow}$ are not normalised; the norm of $\Theta_{\downarrow}$ is much smaller than the one of $\Theta_{\uparrow}$ since the probability of the particle to emerge from the box is much smaller than the probability to remain inside.) Here we are interested in the case when the particle is released. The released particle and opener state is approximatively (see Suplementary Information 2)

\beqa&&
\Theta_{\uparrow}(q,x)=\int e^{-i {{\alpha^2}\over{2}} T}\phi (q-T)e^{-{i}
{{({{k^2}\over2}-{{\alpha^2}\over2})}}q}h(k)e^{ikx}dk,\nonumber\\&&\label{emitted_main}\eeqa
with

\beq h(k)={1\over{2i}}\int_{-L}^L dx' (e^{i(\alpha
- k )x'}-e^{i(-\alpha
- k )x'}).\label{truncatedwave}\eeq

An essential thing to note about Eq. (\ref{emitted_main}) is that it is identical (up to normalisation) to what we would have obtained had we started with the particle in the energy eigenstate  $\sin(\alpha x)$ instead of the ``fake" state $\psi(x)$ of eq. (\ref{particle state}).

To see the meaning of Eq. (\ref{emitted_main}) we first note that since the particle is now free the energy eigenstates are the plane waves $e^{ikx}$. Hence, as far as the particle is concerned, Eq. (\ref{emitted_main}) is actually the decomposition of the state into energy eigenstates. Everything else being phase factors, the probability of the released particle to have momentum $k$ corresponding to energy $k^2/2$ is $|h(k)|^2$. But all we have in (\ref{truncatedwave}) are two truncated wave-trains, corresponding to the (untruncated) plane waves  $e^{\pm i\alpha x}$.  Thus
the most probable final plane-wave state is a free particle in a superposition of eigenstates of momentum
$\pm\alpha$ and energy ${{\alpha^2}\over{2}}$. The particle could also emerge with other energies ${{k^2}\over{2}}\neq{{\alpha^{2}}\over{2}}$ but with smaller probability. The possibility of coming out with these energies is due to the truncation of the wave-train to $|x|\leq L$. Taking $N$ larger, we can make $L$ larger and thus decrease these probabilities, relative to the probability of having energy ${{\alpha^2}/2}$, as much as we want. As noted before, this is exactly what would have happened had we started in the high-energy eigenstate $\sin(\alpha x)$.

One's natural suspicion is that when the particle emerges from the box with energy ${{\alpha^2}/ {2}}$, which is much larger than the maximal energy it had inside the box, the additional energy of the emitted particle comes from the opener.  But the energy distribution of the opener before and after the interaction is the same; $\phi (q)$ is
merely displaced, as if there had been no interaction. This is the paradox.

Note also that when the particle emerges with an energy slightly different from ${{\alpha^2}/{2}}$, because of the truncation of the wave-packet, the opener supplies this small difference (mathematically expressed by the $q$-dependent phase accumulated by the opener) but not the difference between the true low energies that the particle originally had and the high energy with which it emerges.

\section{The standard energy conservation}

As we discussed before, for any time independent Hamiltonian, energy is {\it always} conserved in the standard sense, that is, the probability distribution of the total energy is time independent. Since this is a theorem, it holds in our case as well; our paradox appears only at the level of individual cases. Yet it is worth looking in more detail at the standard account of energy conservation as it applies in our case. As our case has interesting characteristics, the standard account of energy conservation turns out to be interesting as well.

The total Hamiltonian is
\begin{equation}
H = H_{\gamma}+H_{\Omega}+H_{int}
\end{equation}
where $ H_{\gamma}$ and $H_{\Omega}$ represent the free Hamiltonians of the photon and opener respectively, and $H_{int}$ is the interaction Hamiltonian. Energy conservation strictly speaking refers to the distribution of the eigenvalues $E$ of $H$, which includes the interaction term. However, although the interaction Hamiltonian is present at all times, the system is prepared such that the photon and the opener interact only for a finite time.  Indeed, they evolve essentially free, then interact for a finite time and then continue the free evolution. Hence, long before the interaction and long afterwards, we can ignore $H_{int}$ and say that the probability distribution of the free Hamiltonian, $H_{\gamma}+H_\Omega$, is conserved. That is, the distributions of the total energies $H_\gamma +H_\Omega$ long before the interaction, and long after, must be the same.

The total energy long before the interaction and long after it is simply the sum of the free energies, $E = E_{\gamma} + E_\Omega$. With this definition, the standard energy conservation relation is \beq P^{tot}_i(E)=P^{tot}_f(E)\eeq where $P^{tot}$ denotes the probability distribution of the total energy and the indices $i$ and $f$ stand for ``initial" and ``final". Note that all the probability distributions discussed here are over the entire ensemble, including both the cases in which the photon emerged out of the box and the cases when it didn't.

Before the interaction, there is no correlation between the energies of the photon and opener, i.e. the initial joint probability $P^{\gamma,\Omega}_i(E_p, E_\Omega)$ of their energies is the product of their respective individual probability distributions

\begin{equation}
P_{i}^{\gamma,\Omega}(E_p E_\Omega)=P_{i}^\gamma(E_\gamma)P_i^\Omega(E_\Omega).
\label{initial_energy_distribution}\end{equation}

Correspondingly, the initial total energy distribution is given by
\begin{equation}
P^{tot}_i(E)=\int P^\gamma_i(E')P^\Omega_i(E-E') dE'.
\label{convolution}
\end{equation}

After the interaction it is again the case that the probability distributions of the photon and opener are uncorrelated. More precisely,  as discussed in the previous sections, the truncation of the emerging wave packet leads to some correlations between the energy of the photon and the opener (both when the photon emerges out of the box as well as when it remains in the box), but these correlations can be made as small as we want by taking the length $2\pi N$ of the box long enough. Hence

\begin{equation}
P_{f}^{{\gamma},\Omega}(E_{\gamma} E_\Omega)=P_{f}^{\gamma}(E_{\gamma})P_f^\Omega(E_\Omega)
\label{initial_energy_distribution}\end{equation}
and the final total energy distribution is
\begin{equation}
P^{tot}_f(E)=\int P^{\gamma}_f(E')P^\Omega_f(E-E') dE'.
\label{convolution2}
\end{equation}

The main feature of our experiment is that the energy distribution of the opener doesn't change. Indeed, this feature remains also when we look separately at the cases in which the photon remains in the box and the ones in which the photon is emitted. The photon remains in the box if originally it was far from the central region; in this case it doesn't hit the mirror, so the mirror (and its entire moving mechanism) doesn't change energy. On the other hand, when the photon is in the central region it collides with the mirror. However, as we emphasized when we analyzed the paradox, the mirror energy distribution can't change its energy distribution because, by causality, it must act identically to the case in which a true high-energy photon was in the box; just as the high-energy photon just emerges without changing its energy, and thus leaving the mirror's energy unchanged, so must the fake high-energy photon.
Hence

\beq P^\Omega_i(E)=P^\Omega_f(E).\eeq

On the other hand, the final probability distribution for the energy of the particle is different from the initial distribution, $P_f^{\gamma} \neq P_i^{\gamma}$. Indeed, initially the particle was in a superposition of energy eigenstates all smaller than 1. (See Fig. 2.) After the interaction, however, there are cases when the photon emerges from the box and has high energy, much higher than 1, while in other cases it remains inside the box and has low energies, and the low energies average to something less than the original average in order to conserve the total energy average.

All this leads to a surprising situation: The distribution of the energy of the photon changes without being accompanied by a corresponding change in the distribution of the energy of the opener, yet the distribution of the total energy is conserved.

Although the above situation is surprising, it is mathematically consistent, and it has remarkable implications. Denoting the conserved distributions $P^{tot}_i(E)=P^{tot}_f(E)=P^{tot}(E)$ and $P^\Omega_i(E)=P^\Omega_f(E)=P^\Omega(E)$ we obtain

\beqa P^{tot}(E)&=\int P^{\gamma}_i(E')P^\Omega(E-E') dE'\nonumber\\&=\int P^{\gamma}_f(E')P^\Omega(E-E') dE'.\eeqa

By making the Fourier transform of the convolutions (\ref{convolution}) and (\ref{convolution2}) we obtain

\begin{equation}
\tilde P_{i}^{\gamma}(\tau)\tilde P^\Omega(\tau)=\tilde P_{f}^{\gamma}(\tau)\tilde P^\Omega(\tau)
\label{fourier_convolution}
\end{equation}
where, for each index, $\tilde P(\tau)=\int e^{iE\tau}P(E)dE$ is the Fourier transform of $P(E)$.

Equation (\ref{fourier_convolution}) can have a solution with $\tilde P_i^{\gamma}(\tau)\neq\tilde P_f^{\gamma}(\tau)$ if and only if for some values of $\tau$ the Fourier transform  $\tilde P^{\Omega}(\tau)$ is zero and the changes in $\tilde P_{i}^{\gamma}(\tau)$ are confined to these $\tau$ values.

To understand the significance of the above results, we first note the general meaning of the Fourier transform of the energy distribution. Consider a particle prepared in the state $|\Psi(t)\rangle$ and evolving according to a hamiltonian $H$. Then (see Supplementary Information 3)

\beq \tilde P(\tau)=\langle\Psi(t)|\Psi(t+\tau)\rangle.\eeq
Note that since the hamiltonian  is time independent, $\tilde P(\tau)$ is independent of $t$; indeed
$\langle\Psi(t)|\Psi(t+\tau)\rangle$ is independent of $t$.

In our case, we want the opener-photon interaction to take place only for a finite time: the box must be opened, the mirror inserted, then extracted and the box closed, all before information from remote places in the box can reach the opening. In our explicit model, the opener must thus move from far away, going from an initial state in which there is no interaction to an orthogonal state in which there is interaction and then again to a state with no interaction. This is accomplished by moving through a long sequence of orthogonal states both before and after the interaction (as one can explicitly see in the model). Hence, since the interaction should only take a finite time $T$, it must be the case that the wave function $\phi$ of the opener, as it evolves, must obey
\begin{equation}\langle\phi(t)|\phi(t+\tau)\rangle=0\label{orthogonality}\end{equation}
for any time $\tau>T$. Since (\ref{orthogonality}) is independent of $t$, we can take $t$ to be far in the past, when the opener evolved under its free hamiltonian $H_\Omega$.
Hence, the fact that the interaction takes only a finite amount of time implies that $\tilde P^\Omega(\tau)=0$ for $\tau>T$ and this enables the strange behavior of the energy distributions that characterizes our problem.

Note that the opener has the role of a catalyst: its energy distribution doesn't change, yet, without it, the photon's energy distribution could not change, because the energy of the particle would then be the total energy and changing its distribution would violate the standard energy conservation law.

\section{ Modular energy exchange}

It is interesting to examine further the changes in the energy distributions.
Since neither the total energy distribution nor the opener energy distributions change, it is clear that the average energy of the photon cannot change: Indeed, both before the interaction and after

\begin{equation}
\la H\ra=\la H_{\gamma}\ra+\la H_{\Omega}\ra;
\end{equation}
since $\la H\ra$ and $\la H_{\Omega}\ra$ are constant, so is $\la H_{\gamma}\ra$.

Furthermore, given that both initially and finally the energy distributions of the opener and photon are uncorrelated, and that the distributions of total energy and opener energy do not change, one can easily derive the fact that {\it all} the moments of the photon energy distribution $\la H_{\gamma}^n\ra$ are unchanged.

We thus arrive at another remarkable conclusion: the energy distribution of the photon changes although none of its moments change.

At first it seems that something must be wrong - indeed it is generally assumed that the moments of a distribution completely define it. This however is not so. It is actually perfectly possible for a distribution to change without any of its moments changing. In fact, in quantum mechanics this behavior  characterizes some of the most basic phenomena (such as momentum conservation in the two-slit experiment \cite{aharonov, popescu, aharonov2, qp, short});  and many of the ``mysteries" of quantum mechanics have this mathematical effect at their core. This behavior generally stems from deep reasons connected with causality and non-locality - as our present example illustrates.

So, if none of the moments of the photon's energy distribution change, what changes? It is the average of observables that we call ``modular energies" \cite{aharonov2, qp}, as we now show.

Consider the operator $e^{iH\tau}$; we call this operator ``modular energy" since it depends only on the energy modulo ${{2\pi}/{\tau}}$. Each $\tau$ defines a different modular energy.

Since $H$ is a conserved operator, so are its associated modular energies:
\beq \la e^{iH\tau}\ra_i=\la e^{iH\tau}\ra_f.\eeq
As noted in the previous section, long before the interaction and long afterwards, we can replace replace the full Hamiltonian by the free part, so that

\beq\la e^{i(H_{\gamma}+H_{\Omega})\tau}\ra_i=\la e^{i(H_p+H_{\Omega})\tau}\ra_f.\label{mod_free_energy}\eeq
Since there are no correlations between the states of the photon and the opener either before the interaction or after it, from (\ref{mod_free_energy}) we obtain that
\beq\la e^{iH_{\gamma}\tau}\ra_i \la e^{iH_{\Omega}\tau}\ra_i=\la e^{iH_{\gamma}\tau}\ra_f  \la e^{iH_{\Omega}\tau}\ra_f.\eeq

And since the energy distribution of the opener doesn't change, the averages of its modular energy for every $\tau$ don't change either, i.e. $\la e^{iH_{\Omega}\tau}\ra_i=\la e^{iH_{\Omega}\tau}\ra_f$. Hence, the only changes in the energy distributions of the photon are those of averages of the modular energy
corresponding to those values of $\tau$ for which the average of the corresponding modular energy of the opener is zero. (This statement
is just Eq. (\ref{orthogonality}) stated differently.) Here in particular we have

\begin{equation}
\la e^{iH_{\Omega}\tau}\ra=0
\end{equation} for every
$\tau>T$. This means that for  $\tau>T$ the modular energy of the photon may change.

Note the interesting way in which the conservation of modular energy works. The total modular energy is conserved, so if one of two interacting systems changes its modular energy this must be accompanied by changes in the modular energy of the other, yet the average modular energy of the photon (for some $\tau$) changes while the corresponding modular energy of the opener doesn't. This is possible due to the fact that, as opposite to energy, which is an {\it additive} conserved quantity, its modular part is non-additive but multiplicative, and the modular energy of one of the systems, namely the opener, is {\it completely} uncertain, which makes its average zero.

In concluding this section we would like to emphasize that the whole issue of
exchange of modular energy (or momentum) {\it without any
(significant) exchange of any of its moments} (or where the exchange of the moments plays a trivial role) is a general
characteristic of phenomena in which a {\it localized} probe
interacts with a system in an {\it extended} wave function. At the same time, the particular phenomena described in this paper (the high energy of the photon that emerges from the box)  depend on the {\it particular} form (\ref{state}) of the extended wave function. The possibility of exchange of modular
energy without changes in any of the moments of the energy distribution simply opens a window of opportunity, through
which the phenomena described here can manifest themselves. In other words, the exchange of modular energy without changes in the moments of the energy distribution is just a necessary but not a sufficient condition for the phenomena we describe here.

\goodbreak
\bigskip
\centerline{\hbox{ } }
\centerline{\bf \quad IX. DISCUSSION}
\nobreak
To summarize, in our paper we present an effect that raises questions about what we actually mean by conservation laws. The standard approach is statistical and it is good as far as it goes. However, our effect begs the question of what happens in individual cases. A photon,  prepared in a  superposition of low energy states and with no high-energy component whatsoever, comes out of a  box with great energy. It is legitimate to ask where the energy comes from. We showed that the mechanism used for extracting  the photon - the only other system in the problem- did not provide this energy, so we are left with a puzzle.

Our example concerned energy conservation. It is,
however, clear that one can construct examples involving
conservation of other quantities such as momentum
or angular momentum. The phenomenon is therefore a
general one. Thus we argue that the conservation laws of
quantum mechanics must be revisited and extended. Without doing this, we will be missing a large part of the message quantum mechanics is telling us.

\bigskip
\noindent
{\bf Acknowledgements} Y.A. thanks the Israel Science Foundation (Grant No. 1311/14) for support, and also the ICORE Excellence Center �Circle of Light� and the German-Israeli Project Cooperation (DIP). S.P. thanks the ERC AdV Grant NLST. D.R. thanks the John Templeton Foundation (Project ID 43297) and the Israel Science Foundation (Grant No. 1190/13) for support. The opinions expressed in this publication are the authors� and do not necessarily reflect the views of any of the supporting foundations.

\goodbreak
\bigskip

\goodbreak
\bigskip
\centerline{\bf \quad Supplementary Information 1}
\medskip
\nobreak

To compute the time evolution we note that the Schr{\"o}dinger equation of a particle in a box (square potential well) is identical to that of a free particle, except for the constraint that the wave function is zero at the walls (at $x_L$ and $x_R$). Hence, if one finds a free particle solution that happens to be zero at all times at $x_L$ and $x_R$, and at $t=0$ its restriction between $x_L$ and $x_R$ is identical to the initial box wave function, then its restriction between $x_L$ and $x_R$ is also a solution for the particle in the box problem for all times. The wave function (\ref{particle state}) with $-\infty <x<\infty$ is zero at all times at $x_{L,R}=\pm \pi N$ as one can immediately see by expanding the binomials, so it obeys the above conditions. We will thus now solve this free evolution, using the free propagator method.

\bigskip
\noindent
\beq \Psi(x,t)={1\over{\sqrt{2\pi i}}}\int_{-\infty}^{\infty}\Psi(x',0){1\over{\sqrt t}}e^{i{{(x-x')^2}\over{2t}}}dx'.
\eeq
The integral can be computed by using complex integration and changes of contour in the complex plane. Evaluating the first term in the superposition (\ref{particle state}) we obtain
\begin{widetext}

\beqa f(x,t)&=&{1\over{\sqrt{2\pi i}}}\int_{-\infty}^{\infty}\left( {{1+\alpha}\over 2}\exp({i{{x'}\over N}})+{{1-\alpha}\over 2}\exp({-i{{x'}\over N}})\right)^N{1\over{\sqrt t}}\exp({i{{(x-x')^2}\over{2 t}}})dx'\nonumber\\&=&{1\over{\sqrt{2\pi i}}}\int_{-\infty}^{\infty}\left( {{1+\alpha}\over 2} \exp({i{{x'+x}\over N}})+{{1-\alpha}\over 2} \exp({-i{{x'+x}\over N}})\right)^N{1\over{\sqrt t}}\exp({i{{x'^2}\over{2 t}}})dx'
\nonumber\\&=&{1\over{\sqrt{2\pi}}}\int_{-\infty}^{\infty}\left( {{1+\alpha}\over 2}\exp({i{{x'({{1+i}\over{\sqrt2}})+x}\over N}})+{{1-\alpha}\over 2} \exp({-i{{x'({{1+i}\over{\sqrt2}})+x}\over N}})\right)^N{1\over{\sqrt t}}\exp({-{{x'^2}\over{2t}}})dx'.
\eeqa
\end{widetext}
Here the last equality stems from rotating the integral line from the real axis to the $\pi/4$ line.

\bigskip
\noindent
To evaluate the integral, note that the absolute value of the expression in brackets increases as $\exp(x'/{\sqrt2})$ but it is multiplied by $\exp({-{{x'^2}\over{2t}}})$, a term that for large $x'$ decreases faster. For $x'>t $ the later term dominates and hence the contribution to the integral from these values of $x'$ are negligible.

\bigskip
\noindent
Consider now $t\lesssim\sqrt N$. Then the only contributions to the integral come from $x'\lesssim\sqrt N $ so we can expand the exponentials to first order of ${{x'}\over N}$. Furthermore, since we are interested only in the region $x\lesssim\sqrt N $ we can also expand the exponentials to first order of ${{x}\over N}$. We then obtain for $x\lesssim\sqrt N $ and $t\lesssim{{{\sqrt N }}}$

\begin{widetext}
\beqa f(x,t)&=&{1\over{\sqrt{2\pi}}}\int_{-\infty}^{\infty}\left( {{1+\alpha}\over 2}\exp({i{{x'({{1+i}\over{\sqrt2}})+x}\over N}})+{{1-\alpha}\over 2} \exp({-i{{x'({{1+i}\over{\sqrt2}})+x}\over N}})\right)^N{1\over{\sqrt t}}\exp({-{{mx'^2}\over{2t}}})dx'\nonumber\\&\approx&{1\over{\sqrt{2\pi}}}\int_{-\infty}^{\infty}\left( {{1+\alpha}\over 2}(1+{i{{x'({{1+i}\over{\sqrt2}})+x}\over N}})+{{1-\alpha}\over 2} (1{-i{{x'({{1+i}\over{\sqrt2}})+x}\over N}})\right)^N{1\over{\sqrt t}}\exp({-{{mx'^2}\over{2 t}}})dx'\nonumber\\&=&{1\over{\sqrt{2\pi}}}\int_{-\infty}^{\infty}\left( 1+{i\alpha{{x'({{1+i}\over{\sqrt2}})+x}\over N}}\right)^N{1\over{\sqrt t}}\exp({-{{mx'^2}\over{2 t}}})dx'\nonumber\\&\approx&{1\over{\sqrt{2\pi}}}\int_{-\infty}^{\infty} \exp({i\alpha{{x'{{1+i}\over{\sqrt2}}+i\alpha x}}}){1\over{\sqrt t}}\exp({-{{mx'^2}\over{2t}}})dx'=\exp(i\alpha x)\exp(-i{{\alpha^2}\over{2m}}t).
\eeqa

In a similar way we can evaluate the second term of the superposition (\ref{particle state}) and thus obtain that for $x\lesssim\sqrt N $ and $t\lesssim\sqrt N $

\beq\Psi(x,t)\approx\sin(\alpha x)\exp(-i{{\alpha^2}\over{2}}t).\eeq

\end{widetext}

\goodbreak
\bigskip
\centerline{\bf \quad Supplementary Information 2}
\medskip
\nobreak

Consider the following
Hamiltonian to model our experiment:
\begin{equation}{\hat H} =
{{{\hat p}^2}\over{2}}+
V(x) {1+\sigma_z \over 2}+ {\hat p}_q + {{\pi}\over 2}g(x) \delta (q) \sigma_x.\end{equation}
(In this section, to avoid confusions, we will put hats over the hamiltonian and momentum of particle and opener, to indicate that they are operators.)

Let the particle be prepared in the state $\Psi(x,0)=\psi(x)$, with spin $\uparrow$ (i.e. confined inside the box) and the opener in state $\phi (q)$, localised in the region $-T\leq q\leq 0$.

We now calculate the time evolution of the particle-opener system
\beq e^{-i{\hat H}T} \phi (q) \psi (x) |\uparrow \rangle=\Theta_{\uparrow}(q,x)|\uparrow \rangle+\Theta_{\downarrow}(q,x)|\downarrow \rangle.\eeq

Let us first suppose that $\phi(q)=\delta
(q+\tau )$, where $0\le \tau \le T$.  The opener advances from $t=0$, and at $t=\tau$ it
reaches $q=0$.  When it crosses $q=0$, the opener-particle state
acquires a non-abelian phase $e^{ig(x){{\pi}\over 2}\sigma_x}$, as we see by integrating
the Schr\"odinger equation around the time $t=\tau$.  Thereafter, from $t=\tau$ to $t=T$ it evolves
according to the free propagator.  At time $t=T$, therefore, we have the
two-particle state
\begin{equation}\delta (q+\tau -T)
e^{-i{{{\hat H}_0}}(T-\tau )} e^{-ig(x){{\pi}\over 2} \sigma_x} e^{-{i}{{{\hat H}_0}} \tau }
\psi (x) |\uparrow \rangle,\end{equation}
where ${\hat H}_0 = {\hat p}^2/2
+ V(x)(1+\sigma_z )/2$.

The effect of the trigger is to flip the spin in the region $|x|\le L$ where $g(x)=1$ and leave it undisturbed where $g(x)=0$:
\begin{equation}e^{-i{{\pi}\over 2}g(x) \sigma_x} |\uparrow \rangle =
(1-g(x))|\uparrow \rangle-i g(x)|\downarrow \rangle.\end{equation}
In other words, the particle is released if it is in the region $|x|\le L$ and it remains trapped in the box otherwise.

Let us concentrate on the released particles. We obtain

\beqa &&
\Theta_{\downarrow}(q,x)=\nonumber\\&&-\delta (q+\tau -T)
e^{-{i}{{{\hat p}^2}\over{2m}}(T-\tau )} ig(x) e^{-{i}({{{\hat p}^2}\over {2 }}+{{V(x)}})\tau}
\psi (x).\nonumber\\ \eeqa
Note the effect of the spin flip: for time $0<t<\tau$ the particle was confined in the box and evolved according to the Hamltonian ${\hat p}^2/2+V(x)$ and for $\tau<t<T$ the particle was free and evolved according to the Hamltonian ${\hat p}^2/2$.

By construction $g(x)=1$ around the center of the box, i.e. for $|x|<L$ and zero outside this region. As noted before, in this region for times $\tau<T<\sqrt N$ the fake high-energy photon behaves like a true high-energy eigenstate, so we get (up to normalisation)

\beqa &&
\Theta_{\downarrow}(q,x)\approx\nonumber\\&&\delta (q+\tau -T)
e^{-{i}{{{\hat p}^2}\over {2}} (T-\tau )} ig(x)
e^{-{i}{{\alpha^2}\over{2}}\tau }\sin(\alpha x)\label{emitted}.\nonumber\\ \eeqa

A more realistic initial wave function $\phi(q)$ for the opener would have support
over a finite, extended region $-T<q< 0$.  To compute the effect of such a trigger,
we just fold the previous result with this wave function.
We write
\begin{equation}
\phi (q) = \int \phi (-\tau ) \delta (q+\tau )  d \tau
\end{equation}
and we obtain for the released particles

\beqa &&
\Theta_{\downarrow}(q,x)\approx\nonumber\\
 &&\phi (q-T )e^{-{i}{{\alpha^2}\over{2}}T}e^{-{i}({{{\hat p}^2}\over{2}}-{{\alpha^2}\over{2}})q}
g(x)\sin(\alpha x).\nonumber\\
\label{emitted2}\eeqa

An essential thing to note about (\ref{emitted}) and (\ref{emitted2}) is that they are identical (up to normalisation) to what would have happened had we started with the particle in the energy eigenstate  $\sin(\alpha x)$ instead of the ``fake" state $\psi(x)$ of eq. (\ref{particle state}).

To analyze the energy distribution of the emitted particles we write the final photon-opener state using the momentum representation for the particle. Using

\beq g(x)\sin(\alpha x)=\int_{-\infty}^{\infty} h(k) e^{ikx} dk\eeq
with
\beqa h(k)&=&\int_{-\infty}^{\infty} g(x')\sin(\alpha x')e^{-ikx'}dx'\nonumber\\&=&{1\over{2i}}\int_{-L}^L dx' (e^{i(\alpha
- k )x'}-e^{i(-\alpha
- k )x'})\eeqa
and where in the second equality we used the definition of $g(x)$,
we
have (up to normalisation)

\beqa&&
\Theta_{\downarrow}(q,x)\nonumber\\&\approx&\int_{-\infty}^{\infty} \phi (q-T)e^{-{i} {{\alpha^2}\over{2}} T}e^{-{i}
{{({{{\hat p}^2}\over2}-{{\alpha^2}\over2})}}q}h(k)e^{ikx} dk,\nonumber\\&=&\int_{-\infty}^{\infty} \phi (q-T)e^{-{i} {{\alpha^2}\over{2}} T}e^{-{i}
{{({{k^2}\over2}-{{\alpha^2}\over2})}}q}h(k)e^{ikx} dk.\eeqa

\goodbreak
\bigskip
\centerline{\bf \quad Supplementary Information 3}
\medskip
\nobreak

To compute $\tilde P(\tau)$:
\beqa
&\tilde P(\tau)=\int e^{iE\tau}P(E)dE\nonumber\\&=\int e^{iE\tau}\langle\Psi(t)|E\rangle\langle E|\Psi(t)\rangle dE\nonumber\\ &=\langle\Psi(t)|e^{iH\tau}|\Psi(t)\rangle=\langle\Psi(t)|\Psi(t+\tau)\rangle
\eeqa
where $|E\rangle$ denotes an eigenstate of energy and where we used the fact that

\begin{equation}
\int e^{iE\tau} |E\rangle\langle E|dE=e^{iH\tau}.
\end{equation}


\begin{thebibliography}{0}
\bibitem{footnote} This is not contradicted by the existence of the well-known non-local quantum effects such as the Aharonov-Bohm effect \cite{AB} and violations of Bell-inequalities \cite{bell} as well as Aharonov-Bohm type which are local in this sense.
\end{thebibliography}
\end{document}